\documentclass[12pt]{article}
\begin{document}

\begin{center}
{\large \bf Origin of Reactor Antineutrino Anomaly}
\vspace{0.5 cm}

\begin{small}
\renewcommand{\thefootnote}{*}
L.M.Slad\footnote{slad@theory.sinp.msu.ru} \\
{\it Skobeltsyn Institute of Nuclear Physics,
Lomonosov Moscow State University, Moscow 119991, Russia}
\end{small}
\end{center}

\vspace{0.3 cm}

\begin{footnotesize}
\noindent
{\bf Abstract.} The reactor antineutrino anomaly, discovered in 2011, means a noticeable difference between the observed rate of inverse beta decays and the expected (theoretical) rate of such processes based on the measurement of the spectra of electrons in beta decay of $^{235}$U, $^{239}$Pu, $^{241}$Pu, and $^{238}$U nuclei and their subsequent conversion into right-handed antineutrino spectra. This paper provides a rationale for the fact that both right-handed and left-handed antineutrinos are produced in beta decays of nuclei. But the conversion procedure in any case assigns a right-handed antineutrino to each electron, which leads to the superiority of the theoretical rate of inverse beta decays over the experimental one. The right-handed antineutrino is produced in the mode of beta decay of the nucleus due to the standard electroweak interaction. The left-handed antineutrino appears in the mode caused by the existence of a interaction, the carrier of which is a massless pseudoscalar boson having a Yukawa coupling with the electron neutrino and nucleons. The emission of such a boson from a virtual right-handed antineutrino converts it into a free left-handed antineutrino.
\end{footnotesize}

\vspace{0.5 cm}

\begin{small}

\begin{center}
{\large \bf 1. Introduction}   
\end{center}

   A complete study of neutrino physics is not possible without knowledge of the spectra of reactor antineutrinos.

   Finding information about them is implemented in two ways. One of them, which gives the observed electron right-handed antineutrino spectrum, is based on the analysis of inverse beta decays observed in 18 experimental setups (they are listed in Fig. 5 of paper \cite{1}), located at different distances from the reactors. Another way, giving the theoretical spectrum of antineutrinos, was measuring the spectra of electrons from the beta decays of $^{235}$U, $^{239}$Pu, $^{241}$Pu, which was carried out at the ILL reactor in Grenoble, starting in 1979 \cite{2} (electron spectrum induced by $^{238}$U beta decays were found much later at the Garching reactor \cite{3}), and then converting them to electron antineutrino spectra without specifying the beta decay branches \cite{4}-\cite{6}. The refinement of the conversion of electron spectra to antineutrino spectra was performed in papers \cite{7} and \cite{8} by involving thousands of branches of beta decay of uranium and plutonium nuclei into considaration.

   A comparison of the expected rate of inverse beta decays in a reactor setup, found on the basis of the theoretical spectra of reactor antineutrinos, with the observed rates of such processes revealed their noticeable difference. The authors of the work \cite{1} claim that the ratio of the rate of observed events to the rate of expected events is $0.943 \pm 0.023$. The distinction of this ratio from unity is called by them the reactor antineutrino anomaly. The authors believe, without insisting on this, that the named anomaly may indicate the existence of a fourth non-standard neutrino, with $\Delta m^{2}_{\rm new} \geq 1$ eV$^{2}$.

   In a broad review \cite{9}, published 13 years after the reactor antineutrino anomaly was discovered, the following is stated. The observed antineutrino spectra for reactors of various types with various fuel compositions are known with an accuracy of less than $0.5\%$, which is beyond doubt. There may be a defect in the Hubert-Muller model, either in the original ILL $\beta$-spectra data, or in the conversion procedure itself. Sterile neitrino remains a fascinating possibility. 

\begin{center}
{\large \bf 2. The futility of the concept of electron neutrino oscillations}   
\end{center}

   This section is based entirely on the paper \cite{10}, which presents a logical and numerical analysis of all stages in the development of the particle oscillation concept. The total of this analysis is formulated in the above title. 
  
   The particle oscillation concept, originally embodied by Pais and Piccioni \cite{11}, arose as a result of the division of particles, proposed by Gell-Mann and Pais \cite{12}, into true particles and their "mixture", which respectively have and do not have certain masses. Only "mixtures of particles" oscillate. Since in Davis's first experiments \cite{13}, contrary to expectations, not a single transition of chlorine into argon under the influence of solar neutrinos was detected, Gribov and Pontecorvo suggested \cite{14} that the problem of solar neutrinos can be solvable based on the assumption about the existence of neutrino oscillations. Implementing the Gell-Mann-Pais-Piccioni concept, Gribov and Pontecorvo give to the well-known neutrinos $\nu_{e}$ and $\nu_{\mu}$ the status of a "mixture" of new neutrinos $\nu_{1}$ and $\nu_ {2}$ with masses $m_{1}$ and $m_{2}$. However, as argued in the paper \cite {10}, the postulation of the concept of particle oscillation and, in particular, solar neutrino oscillations contradicts the provisions of classical logic. It was also proved there that the fulfillment of the law of conservation of energy-momentum in the process of the production of an electron neutrino is impossible if it is a mixture of new neutrinos whose masses satisfy the relation
\begin{equation}
|m_{1}^{2}-m_{2}^{2}| >  5.9 \cdot 10^{-13} {\rm eV}^{2}.
\label{1}
\end{equation}

   The law of conservation of energy-momentum imposes, in particular, an absolute ban on the existence of a sterille neutrino, the assumption of which is expressed in the work \cite{1} and still remains today as can be seen from the article \cite{9}.

\begin{center}
{\large \bf 3. Interaction carried by massless pseudoscalar bosons.}
\end{center}

   Relatively recently the problem of solar neutrinos received a logically simple and clear solution within the framework of classical methods of quantum field theory \cite{15}. It is based on the hypothesis of the existence of a new interaction, the carrier of which is a massless pseudoscalar boson, having Yukawa couplings with an electron neutrino, proton, and neutron, described by the following relativistically invariant Lagrangian
\begin{equation}
 L = ig_{\nu_{e}ps}\bar{\nu}_{e}\gamma^{5}\nu_{e}\varphi_{ps}+
ig_{Nps}\bar{p}\gamma^{5}p\varphi_{ps}-ig_{Nps}\bar{n}\gamma^{5}n\varphi_{ps},
\label{2}
\end{equation}
and not coupled with the electron at the tree level.

   The masslessness of the pseudoscalar boson makes it somewhat similar to the photon. First, it is stable, and second, it allows processes with its emission similar to processes with photon emission.

   The neutrino present in the Lagrangian (\ref{2}) is described by the bispinor representation of the proper Lorentz group and obeys the Dirac equation. Two solutions with positive energy of the massless free Dirac equation, left-handed and right-handed, describe various states of the same neutrino. Due to the Lorentzian structure of the pseudoscalar current,   
\begin{equation}
\bar{\psi} (p_{2}) \gamma^{5} \psi (p_{1}) = \bar{\psi}_{R} (p_{2}) \gamma^{5} \psi_{L} (p_{1}) - \bar{\psi}_{L} (p_{2}) \gamma^{5} \psi_{R} (p_{1}),
\label{A}
\end{equation} 
the emission of a real or virtual massless pseudoscalar boson $\varphi_{ps}$ causes a change in the handedness of the neutrino.

   Thanks to the interaction (\ref{2}), the neutrino collides with the nucleons of the Sun and, due to the current structure (\ref{A}), at each collision it changes its handedness from left to right and vice versa and loses some of its initial energy. In the works \cite{15}, \cite{16}, the consequences of the Brownian motion of neutrinos in the Sun are described in two ways, complementary to each other: by introducing the notion of the effective number of collisions \cite{15} and by taking the geometric distribution of the number of collisions as a test \cite{16}. Both methods give similar results that are in good agreement with experimental results for the rates of all five observed processes with solar neutrinos: for the nuclear transitions of chlorine into argon and gallium into germanium, for the elastic scattering of neutrinos by electrons, for the disintegration of deuterons caused by charged and neutral neutrino currents. The article \cite{10} summarizes the elegant field-theoretical solution to the solar neutrino problem obtained in the works \cite{15}, \cite{16} and points out the shortcomings of the particle oscillation concept that make it useless in physics.

   In work \cite{15}, the following estimate was obtained for the product of Yukawa coupling constants in the Lagrangian (\ref{2})
\begin{equation}
\frac{g_{\nu_{e}ps}g_{Nps}}{4\pi} = (3.2 \pm 0.2) \cdot 10^{-5}.
\label{3}
\end{equation} 
Each of the constants $g_{\nu_{e}ps}$ and $g_{Nps}$ individually can a priori have values in a fairly large range. Meanwhile, processes in which the massless pseudoscalar boson $\varphi_{ps}$ interacts only with nucleons or only with electron neutrinos present undoubted interest both for the physics of the Sun and for the laboratory physics.

\begin{center}
{\large \bf 4. The regularity of the difference between the observed and theoretical rates of events in reactor experiments}   
\end{center}

   The interaction (\ref{2}) of an electron neutrino with a massless pseudoscalar boson naturally entails the existence of a principled difference between the theoretical and observed rates of events in reactor experiments. The magnitude of this difference is determined by the Yukawa coupling constant of the interaction (\ref{2}) $g_{\nu_{e}ps}^{2}/4\pi$, the estimate of which we still do not have.

  Indeed, the beta decay of nucleus X into nucleus Y leads, due to the standard electoweak interaction, to the production of a right-handed antineutrino
\begin{equation}
{\rm X} \rightarrow {\rm Y} + e^{-} + \bar{\nu}_{R},
\label{4} 
\end{equation}
which is able to cause inverse beta decay. At the same time, the emission of a massless pseudoscalar boson from an antineutrino leads, due to the Lagrangian (\ref{2}), to the transformation of an initially virtual right-handed antineutrino into a free left-handed one
\begin{equation}
{\rm X} \rightarrow {\rm Y} + e^{-} + \bar{\nu}_{L} + \varphi_{ps}, 
\label{5}
\end{equation}
which, within the framework of the standard electoweak interaction, cannot manifest itself, causing the production of a positron. The $\varphi_{ps}$ boson in (\ref{5}) is practically unobservable.

   There is also a mode of beta decay of nuclei with the emission of a $\varphi_{ps}$ boson by one of the nucleons of the X nucleus or the Y nucleus, in which the handedness of the final neutrino is the same as in the main mode. Such a mode influence, and only slightly, the probability of the production of right-handed antineutrinos.
 
   Thus, the existence of interaction (\ref{2}) inevitably leads to that among reactor antineutrinos there are left-handed ones. The question of the production of left-handed antineutrinos in reactors has never been raised in the literature before, and their absence was considered an undoubted fact. That is why, when converting the electron spectrum into the spectrum of reactor antineutrinos, each electron is assigned a right-handed antineutrino. As a result, the theoretical rate of inverse beta decay events exceeds the observed rate of such events due to false right-handed antineutrinos, which are matched to electrons from the mode (\ref{5}) with a left-handed antineutrino.

\begin{center}
{\large \bf 5. Estimation of the Yukawa coupling constant of a massless pseudoscalar boson with a neutrino}   
\end{center}

   In what follows, we limit ourselves to obtaining an estimate of magnitude of the constant $g_{\nu_{e}ps}^{2}/4\pi$ based on consideration of the beta decay of two fission fragments of $^{235}$U (they are named in the work \cite {9}): beta transition of $^{143}$Xe to $^{143}$Cs and beta transition of $^{90}$Br to $^{90}$Kr.

   Since the theoretical calculations of the electron spectra in \cite{7} and \cite{8} concern many thousands of branches of beta decays of uranium and plutonium nuclei, and we are interested only in estimating the value of the constant $g_{\nu_{e}ps}^{2}/4\pi$, it is not practical to consider more than two nuclei.

   We calculate the contribution to the partial and total beta decay rate of nuclei in the modes (\ref{4}) and (\ref{5}) in the following approximations.

   In the amplitudes of the processes (\ref{4}) and (\ref{5}), the nuclei X and Y are associated with Dirac bispinors with masses $M_{X}$ and $M_{Y}$, and the weak charged current of the transition X to Y is taken in option V-A. We neglect the mass of the electron.

   Based on the law of conservation of energy-momentum in beta decay (\ref{4}), we obtain the following relations for the minimum and maximum antineutrino energies, $E_{\nu{\rm min}}$ and $E_{\nu{\rm max}} $, at a given electron energy $E_{e}$:
\begin{equation}
E_{\nu {\rm min}} \equiv H_{1} = \frac{M_{X}^{2}-2M_{X}E_{e}-M_{Y}^{2}}{2M_{X}},
\label{6}
\end{equation}
\begin{equation} 
E_{\nu {\rm max}} \equiv H_{2} = \frac{M_{X}^{2}-2M_{X}E_{e}-M_{Y}^{2}}{2(M_{X}-2E_{e})}.
\label{7}
\end{equation}
The maximum possible value of the electron energy at which $H_{1} = H_{2} = 0$ is equal to
\begin{equation} 
E_{0} = \frac{M_{X}^{2}-M_{Y}^{2}}{2M_{X}}. 
\label{8}
\end{equation}
The maximum value of the difference between the maximum and minimum antineutrino energy as a function of the electron energy $E_{e}$ is as follows:
\begin{equation} 
(H_{2}-H_{1})_{\rm max} = \frac{(M_{X}-M_{Y})^{2}}{2M_{X}}. 
\label{9}
\end{equation}

   For the beta transition of $^{143}$Xe to $^{143}$Cs we have: $M_{X}=133134.478$ MeV, $M_{Y}=133127.232$ MeV, $E_{0}=7.247$ MeV , $(H_{2}-H_{1})_{\rm max}=197$ eV. For the beta transition of $^{90}$Br to $^{90}$Kr we have: $M_{X}=83764.355$ MeV, $M_{Y}=83754.015$ MeV, $E_{0}=10.338$ MeV, $(H_{2}-H_{1})_{\rm max}=638$ eV.

   It follows that when converting the spectrum of electrons into the spectrum of reactor antineutrinos, any of the relations (\ref{6}) or (\ref{7}) can be used with a small error. We rely on relations (\ref{6}) and (\ref{8}) and obtain, as in works \cite{4} - \cite{8}, that
\begin{equation} 
E_{\nu} = E_{0} - E_{e}. 
\label{10}
\end{equation}

  Electron spectra as a function of electron energy, equivalent to the differential probabilities of the beta decay mode with boson emission $\varphi_{ps}$ (\ref{5}) and the main beta decay mode (\ref{4}), are given respectively by the following formulas
$$\frac{d\Gamma_{\varphi}}{dE_{e}} = C \frac{g_{\nu ps}^{2}}{4\pi} \left[ \left[ \frac{1}{4}(M_{X}^{2}-M_{Y}^{2})H_{2}^{2}-\frac{1}{3}M_{X}H_{2}^{3}\right]\ln\left( \frac{H_{2}}{\lambda}\right)  \right. \nonumber$$
$$\left. - \left[ \frac{1}{4}(M_{X}^{2}-M_{Y}^{2})H_{1}^{2}-\frac{1}{3}M_{X}H_{1}^{3}\right]\ln
\left( \frac{H_{1}}{\lambda}\right) \right] + \frac{1}{6} (H_{2}^{4}-H_{1}^{4}) \nonumber$$
\begin{equation}
 - \frac{1}{8}(M_{X}^{2}-M_{Y}^{2})(H_{2}^{2}-H_{1}^{2})+\frac{1}{9} M_{X}(H_{2}^{3}-H_{1}^{3}) \equiv C \frac{g_{\nu ps}^{2}}{4\pi} A_{\varphi}(E_{e}) ,
\label{11}
\end{equation}
\begin{equation}
\frac{d\Gamma_{0}}{dE_{e}} = C 4\pi  \left[ \frac{1}{2} (M_{X}^{2}-M_{Y}^{2})(H_{2}^{2}-H_{1}^{2})-\frac{1}{3}M_{X}(H_{2}^{3}-H_{1}^{3})\right]\equiv C A_{0}(E_{e}) ,
\label{12}
\end{equation}
where $C$ is the common factor in the squares of the matrix elements of both beta decay modes. To eliminate the massless divergence the mass $\lambda$ of the virtual antineutrino is introduced. 

   Taking into account the relation (\ref{10}), we perform calculations for 100 values of antineutrino energy exceeding the inverse beta decay threshold of 1.804 MeV. The constant $\lambda$ is successively taken equal to 1 eV, 0.001 eV and 0.000001 eV. First, we find the numerical values of $A_{\varphi}(E_{e})$ and $A_{0}(E_{e})$ as functions of $E_{e}$, and then the values of these quantities, $A_{\varphi}$ and $A_{0}$, after summation over the antineutrino energy from 1.804 MeV to $E_{0}$. We equate the resulting ratio of the probabilities of the two beta decay modes (\ref{5}) and (\ref{4}) $(g_{\nu_{e}ps}^{2}/4\pi) \cdot (A_{\varphi}/A_{0})$, to the corresponding ratio of the event rates $\Gamma_{\varphi}/\Gamma_{0}$ extracted from the equality given in \cite{9}:
\begin{equation}
\frac{\Gamma_{0}}{\Gamma_{0}+\Gamma_{\varphi}} = 0.943 \pm 0.023 .
\label{13}
\end{equation}

The final results are presented in Table 1.
\begin{center}
\begin{tabular}{lcccc}
\multicolumn{5}{c}{{\bf Table 1.} Consequences of two beta decay modes (\ref{4}) and (\ref{5})} \\
\hline
\multicolumn{1}{l}{} 
&\multicolumn{1}{c}{Beta transition} 
&\multicolumn{1}{c}{$\lambda$}
&\multicolumn{1}{c}{$A_{\varphi}/A_{0}$}
&\multicolumn{1}{c}{$g_{\nu ps}^{2}/4\pi$} \\ 
\hline
&$^{143}{\rm Xe} \rightarrow ^{143}{\rm Cs}$ & 1 eV & 0.329 & $0.182 \pm 0.079$ \\
&$^{143}{\rm Xe} \rightarrow ^{143}{\rm Cs}$ & 0.001 eV & 0.480 & $0.125 \pm 0.054$ \\
&$^{143}{\rm Xe} \rightarrow ^{143}{\rm Cs}$ & 0.000001 eV & 0.630 & $0.095 \pm 0.041$ \\
&$^{90}{\rm Br} \rightarrow ^{90}{\rm Kr}$ & 1 eV & 0.345 & $0.174 \pm 0.075$ \\
&$^{90}{\rm Br} \rightarrow ^{90}{\rm Kr}$ & 0.001 eV & 0.499 & $0.120 \pm 0.052$ \\
&$^{90}{\rm Br} \rightarrow ^{90}{\rm Kr}$ & 0.000001 eV & 0.654 & $0.092 \pm 0.040$ \\
\hline
\end{tabular}
\end{center}

Since it has been experimentally established \cite{17} that the upper limit of the neutrino mass is 0.8 eV, the value of the constant $\lambda$, equal to 1 eV, should be considered unacceptable and it should be focused on values of 0.001 eV and 0.000001 eV. By that, we accept the following estimate of the Yukawa coupling constant of the massless pseudoscalar boson with the neutrino 
\begin{equation}
\frac{g_{\nu ps}^{2}}{4\pi} = 0.11 \pm 0.05 .
\label{14}
\end{equation}

It is evident from Table 1 that for each value of the constant $\lambda$ the values of the Yukawa coupling constant $g_{\nu ps}^{2}/4\pi$ obtained for the two selected branches of beta decay are close to each other. 

Using equalities (\ref{3}) and (\ref{14}), we have the following estimate of the Yukawa coupling constant of a massless pseudoscalar boson with nucleons
\begin{equation}
\frac{g_{N ps}^{2}}{4\pi} = (0.93 \pm 0.54) \cdot 10^{-9}.
\label{15}
\end{equation}

Let us note that when solar neutrinos are produced, there is a probability of emission of free massless pseudoscalar bosons from them. It seems plausible that the flux of such bosons from each source of solar neutrinos is given by a factor of order 0.06 from the corresponding neutrino flux. However, the detection of free bosons $\varphi_{ps}$ by converting them into photons on nucleons is impossible due to the extremely low Yukawa coupling constant of a massless pseudoscalar boson with nucleons (\ref{15}).

\begin{center}
{\large \bf 6. A remark about the gallium anomaly}   
\end{center}

The notion of gallium anomaly, which appeared earlier than the notion of reactor antineutrino anomaly, means the difference between the numbers of observed and expected events of transitions of $^{71}$Ga to $^{71}$Ge under the influence of left-handed electron neutrinos from the decay of $^{51}$Cr. At present, there is a scatter in the ratio of these numbers obtained in six variants of completed experiments, in the range from $0.95 \pm 0.12$ to $0.77 \pm 0.05$ \cite{18}, \cite{19}. 

The appearance of the named difference associated with the decay of $^{51}$Cr is inevitable due to the emission of a massless pseudoscalar boson from neutrinos, as a result of which, along with the main electron capture mode
\begin{equation}
^{51}{\rm Cr} + e^{-} \rightarrow ^{51}{\rm V} + \nu_{L}
\label{16}
\end{equation}
there is another mode
\begin{equation}
^{51}{\rm Cr} + e^{-} \rightarrow ^{51}{\rm V} + \nu_{R} +\varphi_{ps}.
\label{17}
\end{equation} 
that cannot lead to transitions of $^{71}$Ga to $^{71}$Ge.

In order to find the level of the gallium anomaly, taking into account the equality (\ref{14}), it is necessary to carry out accurate calculations of the cross sections of both indicated electron capture modes. Apparently, it would not be amiss to re-analyze all the theoretical and experimental aspects of setting up an experiment with $^{51}$Cr.

\begin{center}
{\large \bf 7. Conclusion}   
\end{center}

The discovery in 2011 of the reactor neutrino anomaly is one of the most striking events in neutrino physics. The anomaly appeared as a result of an unconscious substitution of reactor left-handed antineutrinos  with right-handed ones. The anomaly serves as futher evidence of the truth of the hypothesis of the existence of an interaction, the carrier of which is a massless pseudoscalar boson that has a Yukawa coupling with the electron neutrino and nucleons.

\end{small}
\end{document}